\documentstyle[12pt]{article}
\begin{document}
\begin{center}
{\Large \bf PARTICLE CONTENT IN TOPOLOGICAL FIELD THEORIES\footnote{Ref:
OUTP-92-30P}}\\[1.5cm]
{\large Andrew Toon\footnote{Supported by the Royal Society, England.}\\
University of Oxford\\
Department of Theoretical Physics\\
1 Keble Road\\
Oxford, OX1 3NP\\
England}
\end{center}
\vspace*{1in}
\begin{abstract}
By demanding that the path integral measure of topological field theories be
metric independent, we can derive powerful constraints on the particle content
of a topological field theory as well as on the dimensionality of space-time.
\end{abstract}
\newpage
\section{Introduction}
Anomalies impose powerful constraints on quantum field theories. The well known
examples being the necessity of the top quark in the standard model to cancel
the triangle anomaly, and the critical dimensions of string theory to cancel
the conformal anomaly. Both these examples illustrating the power of anomalies
on the particle content of a field theory.

Amomalies are also expected to arise in topological quantum field theories.
Arise in the sense that we wish to preserve topological invariance of these
theories in the quantum theory which is manifest in the classical theory. The
usual argument that shows topological invariance of, in particular, the
partition function goes as follows [1]. Consider any general coordinate
invariant field theory on some $n$-dimensional manifold $M_{n}$ with some
classical action $S$. The full gauged fixed quantum action $S_{q}$ then takes
the form:
\begin{equation}
S_{q}[\Phi_{r},g_{ij}]=S[\Phi_{r}]+\delta_{Q}V[\Phi_{r},g_{ij}].
\end{equation}
Here, $\Phi_{r}$ ($r=1,2,...$) are the fields of the theory including matter,
gauge, ghost and auxiliary fields etc. The second term on the right hand side
of equation (1) is the ghost plus gauge fixing term which is associated with
all the gauge invariance of $S[\Phi_{r}]$. Note that the metric $g_{ij}$ on
$M_{n}$ only appears in this term, that is one introduces the metric $g_{ij}$
only in the process of gauge fixing. $\delta_{Q}$ stands for the related
nilpotent BRS-transformation and thus the entire gauge fixing plus ghost term
is written as a BRST variation of some functional $V[\Phi_{r},g_{ij}]$.
Naively, the partition function is then given by:
\begin{equation}
Z(M_{n},g_{ij})=\int D[\Phi]\exp iS_{q}[\Phi_{r},g_{ij}].
\end{equation}
Following Witten's argument, topological invariance is preserved in the quantum
theory since [1]:
$$\frac{\delta}{\delta g^{ij}}Z(M_{n},g_{ij})=\int D[\Phi] \exp
(iS_{q})\frac{\delta}{\delta g^{ij}}(\delta_{Q}V)$$
\begin{equation}
=\int D[\Phi] \exp(iS_{q})\delta_{Q}(\frac{\delta}{\delta g^{ij}}V)=0,
\end{equation}
by BRST invariance. What is crucial in the above derivation is the assumption
that the path integral measure $D[\Phi]$ is metric independent. Once this is
established, it is clear from the above argument that topological invariance is
preserved in the quantum theory. That is the partition function of the theory
will be metric independent. However, as we will show, there exists a large
class of ^^ ^^ topological field theories" which have a metric dependent
partition function but whose vacuum expectation value of some metric
independent  BRST invariant operator is topologically invariant.

The outline of this paper is as follows. In section 2, we will study the path
integral measure in topological field theories and derive a constraint that
preserves topological invariance in the quantum theory. We then use this
constraint to investigate the particle content of topological field theories
which have  topologically invariant partition functions. We then discuss a
large class of ^^ ^^ topological field theories" which have metric dependent
partition functions but whose vacuum expectation value of some metric
independent BRST invariant operator is automatically metric independent.
\section{Path integral measure in topological field theories}
We will now show that in general the path integral measure of topological field
theories is not metric independent. We will use the method of reference [2] who
in turn expliot the method of Fujikawa [3].

Consider any general coordinate invariant field theory in $n$-dimensions with
quantum action $S_{q}$ given by equation (1). Recall that when one gauge fixes
this theory, one picks a metric $g_{ij}$ on $M_{n}$. The partition function is
then not the naive partition function of equation (2) but is given by:
\begin{equation}
\tilde{Z}(M_{n},g_{ij})=\int\tilde{D}[\Phi]\exp iS_{q}[\Phi_{r},g_{ij}],
\end{equation}
where $\tilde{D}[\Phi]$ stands for an appropriate general coordinate invariant
measure over the full set of fields $\Phi_{r}$. Fujikawa [3] as proposed the
following choice of the measure for being general coordinate invariant:
\begin{equation}
\tilde{D}[\Phi]=\prod_{x}[\prod_{r}d\tilde{\Phi}_{r}(x)],
\end{equation}
where for any given field component $\Phi_{r}$:
\begin{equation}
\tilde{\Phi}_{r}(x)=g^{\alpha_{r}}(x)\Phi_{r}(x),\; g(x)=\det g_{ij}.
\end{equation}
The $\alpha_{r}$ are constants which depend upon the tensor nature of the
fields as well as the space-time dimension. Some examples being:
\begin{equation}
\alpha_{r}=\left\{ \begin{array}{ll}
\frac{1}{4} & \mbox{for each scalar field}\\
\frac{n-2}{4n} & \mbox{for each component of a covariant vector field}\\
\frac{n-4}{4n} & \mbox{for each component of a tensor field of rank two.}
\end{array}
\right.
\end{equation}
It is now clear that:
\begin{equation}
\prod_{x}d\tilde{\Phi}_{r}(x)=\prod_{x}d[g^{\alpha_{r}}(x)\Phi_{r}(x)]=\prod_{x}[g^{\alpha_{r}\sigma_{r}}(x)d\Phi_{r}(x)],
\end{equation}
where the signature $\sigma_{r}$ is +1 $(-1)$ for commuting (anti-commuting)
fields. Thus, the general covariant Fujikawa measure becomes:
\begin{equation}
\tilde{D}[\Phi]=\prod_{x}[g(x)]^{K}(\prod_{r,y}d\Phi_{r}(y)),
\end{equation}
where:
\begin{equation}
K=\sum_{r}\sigma_{r}\alpha_{r},
\end{equation}
is an index which measures the metric dependence of the path integral measure.
The partition function (4) thus becomes:
\begin{equation}
\tilde{Z}(M_{n})=(\prod_{x}[g(x)]^{K})Z(M_{n}),
\end{equation}
where $Z(M_{n})$ is the partition function using the naive measure
$\prod_{x,r}[d\Phi_{r}(x)]$. Thus for topological invariance to be preserved at
the quantum level we require:
\begin{equation}
K=\sum_{r}\sigma_{r}\alpha_{r}=0.
\end{equation}
\section{Particle content in topological field theories}
We saw in the previous section that in order to preserve topological invariance
of a classical topological theory in the quantum theory requires that $K=0$.
Note also that for supersymmetric type theories that $K\equiv 0$ and thus
equation (12) gives no useful information about these types of theories.
Excluding such theories we will now use equation (12) to derive the particle
content of topological field theories in various dimensions.

The simplest theory that we would like to write down is a gauge theory, with
some gauge group $G$, together with a Lagrange multiplier which enforces some
gauge constraint, ghost and anti-ghost fields. In n-dimensions, we then
require:
\begin{equation}
K=\sum_{r}\sigma_{r}\alpha_{r}=dim(G)(n.\frac{(n-2)}{4n}+\frac{1}{4}-\frac{1}{4}-\frac{1}{4})=0.
\end{equation}
Solving singles out $n=3$ which corresponds to the Chern-Simons theory in 3
dimensions which has classical action:
\begin{equation}
S=\frac{1}{2}\int_{M_{3}}<A\wedge dA+\frac{2}{3}A\wedge A\wedge A>,
\end{equation}
where $A$ is a Lie algebra valued one form and $<\; >$ denotes the trace over
the gauge indices. It is interesting to note that the simplest topological
gauge theory singles out the most complicated space topologically.

In view of the relation of the Chern-Simons theory to topological gravity let
us for the moment stay in 3 dimensions and ask if one can couple topological
matter to this theory with out ruining the topological invariance in the
quantum theory. Consider then a scalar field which will contribute 1/4 to $K$.
To cancel
this we need to couple it to a quantity that contributes $-1/4$ to $K$. This is
achieved by coupling it to a rank two tensor field which contributes
$3\times(3-4)/4.3=-1/4$. This is precisely the topological matter coupling
considered in [4] with topological matter action given by:
\begin{equation}
S=\int_{M_{3}}\epsilon^{ijk}(\partial_{i}\phi) B_{jk},
\end{equation}
with $\phi$ being the scalar field and $B_{ij}$ being a rank two anti-symmetric
tensor field.

Let us now return to $n$ dimensions. Consider again a gauge theory, with some
gauge group $G$, together with a Lagrange multiplier, ghost and anti-ghost but
now coupled to $\gamma$ scalar particles in the adjoint representation of the
gauge group. Equation (12) now gives:
\begin{equation}
K=\sum_{r}\sigma_{r}\alpha_{r}=dim(G)(n.\frac{(n-2)}{4n}+\frac{1}{4}-\frac{1}{4}-\frac{1}{4}+\frac{\gamma}{4})=0.
\end{equation}
Solving gives $n=3-\gamma$. The case $\gamma=1$ giving precisely the particle
content of 1+1 topological gravity with classical action given by [5]:
\begin{equation}
S=\frac{1}{2}\int_{M_{2}}\phi_{A}F_{\alpha\beta}^{A}\epsilon^{\alpha\beta},
\end{equation}
where $\phi_{A}$ is the scalar field in the adjoint representation and
$F_{\alpha\beta}^{A}$ is the field strength of the gauge field. Again, staying
with 2-dimensions, it is interesting to ask if one can couple topological
matter fields to this theory. Again, lets consider a scalar field $\Phi$
(different from $\phi$) which contributes 1/4 to $K$. To cancel this we can
couple it to an anti-symmetric tensor field, $B_{ij}$, of rank two which
contributes $1\times(2-4)/4.2=-1/4$ to $K$. This has classical action:
\begin{equation}
S=\int_{M_{2}}\epsilon^{ij}\Phi B_{ij},
\end{equation}
where it should be clear that $\Phi$ and $B_{ij}$ transform the same way under
the gauge group $G$ being considered. Note also that in [5] matter coupling of
a scalar field to a vector field is considered which would imply $K\neq 0$.
However, what happens under this coupling is that the symmetry of the theory is
increased resulting in the need for further gauge fixing. It is precisely this
further gauge fixing that restores $K=0$.

\section{Pseudo-topological field theories}
The previous section concerned topological field theories whose partition
function turned out to be topological invariants. However, there exists a large
class of topological field theories whose partition function is not topological
invariant but whose vacuum expectation value of some gauge invariant operator
is. Consider for example the following theories on some odd dimensional
manifold $M_{2n+1}$ with classical action given by [6]:
\begin{equation}
S=\int_{M_{2n+1}}\Omega_{2n+1},
\end{equation}
where $\Omega_{2n+1}$ is the generalised Chern-Simons $2n+1$-form:
\begin{equation}
\Omega_{2n+1}=(n+1)\int_{0}^{1}dt<A(tdA+t^{2}A^{2})^{n}>,
\end{equation}
with $A$ being a Lie algebra valued one-form. Since, just like the Chern-Simons
3-form, the classical action (19) is indepenent of any metric on $M_{2n+1}$,
one would expect to obtain only topological information from these field
theories. However, to quantise this theory we note
that we have a single gauge field and thus we need to introduce a Lagrange
multiplier to gauge fix plus the corresponding ghosts and anti-ghosts. Using
equation (12) we see that $K=0$ only in 3-dimensions. Thus, it appears that the
generalised Chern-Simons theory will have a ^^ ^^ topological anomaly" for all
odd
dimensions different from 3. There also appears to be a corresponding problem
with the even dimensional models considered in [6] where a single scalar field
is coupled to  the field strength of some gauge field. We see this from
equation (12) which seems to imply a ^^ ^^ topological anomaly" in all even
dimensions different from 2.

The above models may not in general have topological invariant partition
functions but their vacuum expectation value of some gauge invariant quantity
will for the following reasons. We now assume that $K\neq 0$ and consider the
vacuum expectation value of any classically topologically invariant quantity
$W[\Phi_{r}]$. We thus have:
\begin{equation}
\tilde{<W>}=\frac{\int \tilde{D}[\Phi]W\exp iS_{q}}{\int \tilde{D}[\Phi]\exp
iS_{q}}=\frac{[\prod_{x}g(x)]^{K}\int D[\Phi] W\exp
iS_{q}}{[\prod_{x}g(x)]^{K}\int D[\Phi]\exp iS_{q}}=<W>,
\end{equation}
where $<W>$ is the vacuum expectation value of $W$ using the naive measure
$\prod_{x,r}[d\Phi_{r}(x)]$. Thus, for non zero $K$, we see that the metric
dependence cancels giving a metric independent result which follows from the
argument given by equation (3). Thus if $K\neq 0$ we see that one may still
obtain topological information even though the partition function is not a
topological invariant. See for example reference [7] where the 5-dimensional
generalised Chern-Simons term is discussed and its relation to link invariants.
\section{Conclusion}
We have seen that by demanding the path integral measure in topological field
theories be metric independent, we are led to powerful constraints on the
particle content of a theory as well as on the dimensionality of space-time.
These constraints being particularly powerful if we want the theories partition
function to be a topological invariant.

As we have seen, given a classical action which is metric independent, we can
have two types of topological field theories. One with $K=0$ and the other with
$K\neq 0$, which we called pseudo-topological field theories because their
partition function turns out to be metric dependent but their vacuum
expectation value of some gauge invariant operator is not. One is now led to
the question wheather the $K\neq 0$ models might have possible applications to
physics.
This being a natural question since the main problem of taking topological
field theories seriously as physical models is how can some metric structure
arise in these theories?

The general idea of how topological field theories may find application to
physics goes as follows. One imagines that the initial state of the universe is
in an unbroken phase. That is the metric of space-time is zero and thus there
being no notion of physics as we know it. It is this phase that is described by
a
topological field theory. Once this view is accepted, the most fundamental
problem becomes why is the space-time metric we see today nondegenerate instead
of
zero? One would expect some kind of Higg's mechanism to come to our rescue here
as was done in [8,9] but to do this one must impose some hidden structure
on the unbroken phase of the theory. That is, for example [9], start of with
matter fields already in the theory, show that the theory admits an unbroken
phase (show that the unbroken phase is a solution to the classical equations of
motion) and then argue that a possible vacuum instability may occur leading to
nondegenerate metrics. The problem then with a Higg's like mechanism is that
one must start with a theory that already has physical structure but admits a
diffeomorphism invariant phase. This really running counter with the notion of
topological field theories since the initial phase they are supposed to
describe has no metric structure and thus no physics. It really describes a
dead or unborn universe.

The pseudo-topological field theories, therefore, might have a natural solution
to the above problem. That is, the unbroken phase of gravity being described by
the vacuum expectation value of some operator which we know to be metric
independent. The broken phase then being the simplest ^^ ^^ observable" in the
theory, namely the partition function which is metric dependent. This idea at
least
confirming many beliefs that physics should be simple.

This research was supported by the Royal Society England.
\newpage
\begin{center}
\bf{References}
\end{center}

[1] E. Witten, Commun. Math. Phys. 117 (1988) 353.\\

[2] R. Kaul, R. Rajaraman, Phys. Lett. B249 (1990) 433.\\

[3] K. Fujikawa, Phys. Rev. Lett. 42 (1979) 1195\\

[4] J. Gegenberg, G. Kunstatter, H. Leivo, Phys. Lett. B252 (1990) 381.\\

[5] A. Chamseddine, D. Wyler, Nucl. Phys. B340 (1990) 595.\\

[6] A. Chamseddine, Nucl. Phys. B346 (1990) 213.\\

[7] P. Gusin, Mod. Phys. Lett. A, Vol. 7, No. 34 (1992) 3203.\\

[8] R. Percacci, Nucl. Phys. B353 (1991) 271.\\

[9] S. Giddings, Phys. Lett. B268 (1991) 17.
\end{document}